\documentclass[aps,prb,twocolumn,showpacs,superscriptaddress]{revtex4-1}
\usepackage{graphicx}
\usepackage{color}

\begin{document}

%\preprint{}

\title{How chromium doping affects the correlated electronic structure of V$_2$O$_3$}

\author{Daniel Grieger}
%\email[]{dgrieger@sissa.it}
%\thanks{}
\affiliation{SISSA, Via~Bonomea~265, I-34136~Trieste, Italy}

\author{Frank Lechermann}
%\thanks{}
\affiliation{I. Institut f\"ur Theoretische Physik,
  Universit\"at~Hamburg, Jungiusstr.~9, D-20355~Hamburg, Germany}

\date{\today}

\begin{abstract}
  The archetypical strongly correlated Mott-phenomena compound V$_2$O$_3$
  is known to show a paramagnetic metal-insulator transition 
  driven by doping with chromium atoms and/or (negative) pressure. Via
  charge self-consistent density-functional theory+dynamical mean-field theory
  calculations we demonstrate that these two routes cannot be understood as equivalent. 
  To this end, the explicit description of Cr-doped V$_2$O$_3$ by means of supercell 
  calculations and the virtual crystal approximation is performed. Already the
  sole introduction of chromium's additional electron to the system 
  is shown to modify the overall correlated electronic structure substantially. 
  Correlation-induced charge transfers between Cr and the remaining V ions occur and the 
  transition-metal orbital polarization is increased by the electron doping, in close 
  agreement with experimental findings.
\end{abstract}

%\keywords{}

\maketitle

\section{Introduction}
The vanadium sesquioxide V$_2$O$_3$ is among the most prominent
strongly correlated compounds and has already been studied in many
theoretical
works.~\cite{cas78,mat94,ezh99,elf03,eye05,hel01,kel04,pot07,san13,guo14,gri12}
The three major phases found in its prototype phase
diagram~\cite{mcw71,mcw73} at temperature $T$ are a paramagnetic metal
(PM) and a paramagnetic insulator (PI) based on the corundum crystal
structure as well as a monoclinic antiferromagnetic insulating phase
at lower $T$. Note that this picture might not even be complete, as
the discovery of new more exotic phases such as paramagnetic
monoclinic~\cite{din14} or spin density wave~\cite{bao93} structres
suggests.

 At ambient $T$ and pressure $p$, the stoichiometric compound is in the metallic 
state. Along the crystallographic $c$-axis of the corundum structure V-V pairs appear
and a honeycomb lattice marks the $ab$-plane. The V ions reside within an octahedron of 
O ions, which in the present case build up a trigonal crystal field around the 
transition metal. The low-energy $t_{2g}$ orbitals of the V($3d$) shell are split into 
an $a_{1g}$ and two degenerate $e_g'$ orbitals. Formally the vanadium ion has the $3d^2$ 
valence configuration, i.e. is in the V$^{3+}$ oxidation state. 

In experiment the traditional phase changes at constant temperature are realized through 
doping the material either with Cr or Ti. While chromium doping allows for the 
metal-to-insulator transition (MIT), Ti-doping is more relevant to map out phase 
regions at rather high/low temperature. Since the readily measurable structural effect of 
adding Cr lies in some lattice expansion,~\cite{mcw71,mcw73} the role of Cr-doping in 
literature is often fully reduced to the picture of applying negative $p$, therewith rendering 
the V$_2$O$_3$ MIT to a realistic example of Mott's original thinking
about electron localization. However substitutional Cr is not isovalent with V, but
has one additional electron in the valence. Yet the details of this specific electron 
doping as well as the resulting Cr-induced electronic-structure effects have so far not 
really been studied on an elaborate level. In other words, the explicit defect-related 
physics of Cr-doping in V$_2$O$_3$ may bear much more processes relevant for the driving 
forces behind the Mott transition than pure structural effects. In this regard the
elder orbital-polarization measurements of Park {\sl et al.}~\cite{par00} are especially 
interesting. They revealed a a much larger vanadium $e_g'/a_{1g}$ orbital polarization in 
the (Cr-doped) PI than in the metallic phase. Recenty 
Rodolakis {\sl et al.}~\cite{rod10,rod11} compared the Cr doping-driven 
MIT with a truly pressure-driven MIT, i.e. by increasing pressure on insulating 
(V$_{0.92}$Cr$_{0.08}$)$_2$O$_3$. It was observed that in the latter case the
($e_g',a_{1g}$) orbital occupations hardly vary across the MIT. Hence the enhanced 
orbital polarization in the PI appears to be bound to the explicit
doping-driven character of the MIT realization. This observation is confirmed by recent 
calculations based on the charge self-consistent combination of density functional theory (DFT)
with dynamical mean-field theory (DMFT), the so-called DFT+DMFT approach.~\cite{gri12} 
There lattice-expanded stoichiometric V$_2$O$_3$ became Mott-insulating, but indeed 
without a strong change in the $e_g'/a_{1g}$ orbital polarization.

In this work we explicitly show that the Cr doping of V$_2$O$_3$ can be understood as a key constituent of the increased $e_g'/a_{1g}$ orbital polarization. Charge
self-consistent DFT+DMFT applied to substitutional defect supercells as well as in
the virtual crystal approximation (VCA) reveal that the additional valence electron
of chromium is not a passive spectator only giving rise to a change in total volume.
Instead it plays and active part in the correlated electronic structure and 
supposingly appears as a vital ingredient in the mechanisms underlying the V$_2$O$_3$ MIT.

\section{Theoretical framework}
Our first-principles DFT+DMFT approach~\cite{gri12} is based on a
mixed-basis pseudopotential (MBPP) technique~\cite{mbpp_code} for the
DFT part and a hybridization-expansion continous-time quantum
Monte-Carlo impurity scheme~\cite{wer06,triqs_code} to solve the DMFT
impurity problems posed by the realistic materials problem.  The local
density approximation (LDA) to DFT is utilized, crosschecks with the
generalized-gradient approximation (GGA) do not lead to relevant
modifications of the results.  Throughout the work we rely on
correlated subspaces composed of the projected ($e_g',a_{1g}$)
orbital~\cite{ama08,ani05,kar11} threefolds from the Kohn-Sham bands
with dominant transition-metal $3d$ weight. Thus at each charge
self-consistent convergence step a multi-orbital Hubbard Hamiltonian
employing the complete rotational invariant Coulomb interactions is
applied at each correlated site taking part in the overall correlated
subspace. For the parametrization of the Coulomb integrals we choose a
Slater-Kanamori form with Hubbard $U$=5~eV and and Hund's exchange
$J_{\rm H}$=0.93~eV for both Cr and V orbitals, as already utilized in
earlier simplified LDA+DMFT studies for V$_2$O$_3$.~\cite{hel01}

Cr doping of vanadium sesquioxide is computationally realized in two ways. First via the 
replacement of a single vanadium atom by a chromium atom in the original four-V-site 
unit cell as well as in an eight-V-site supercell. Accordingly, this amounts to 
a doping level of $x$=0.25 and $x$=0.125, in our chosen V$_{2(1-x)}$Cr$_{2x}$O$_2$ notation,
respectively. Albeit larger than in most experiments,
these scenarios are geared to corrobate the key qualitative doping effects. Second
we compared this direct approach to a treatment within VCA, allowing for in principle
arbitrary doping levels in an averaged-medium picturing of the problem. There the 
V pseudopotential is replaced by an effective pseudopotential with nuclear charge 
$\tilde{n_c}$=$x\,n_c^{\rm Cr}$$+$$(1$$-$$x)\,n_c^{\rm V}$, with $n_c^{\rm Cr}$,
$n_c^{\rm V}$ as the Cr,~V nuclear charges. Note however that explicit self-energy
effects due to disorder, i.e. finite lifetimes due to a given disorder strength, are
not included in the VCA approach. However it is believed that such direct effects of
disorder are to a good approximation negligible compared to the dominant impact of
Coulomb correlations in V$_2$O$_3$.

In order to include the effect of local structural relaxations due to
Cr doping, the Wyckoff positions of the atomic sites are optimized within 
LDA for all structures used in this work. 
The lattice parameters have been fixed at their values of stoichiometric V$_2$O$_3$ at
ambient pressure in the metallic phase, as found in Ref.~\onlinecite{der70}. Possible 
effects of the enlargement of the unit cell have been published previously in 
Ref.~\onlinecite{gri12}. The $\frac ca$ ratio of the hexagonal lattice 
parameters, which changes in a nontrivial way at the
metal-insulator transition,~\cite{der70} may be expected to have a
significant influence on the electronic structure and is thus
investigated explicitly in section~\ref{caratio}.

\section{LDA perspective of Cr-doped V$_2$O$_3$\label{result1}}
Before tackling the effects of strong electronic correlations in a 
realistic many-body scope, it is useful to take a closer look at how Cr doping
affects the plain LDA electronic structure.
%%%%%%%%%%%%%%%%%%%%%%%%%%%%%%%%%%%%%%%%%%%%%%%%%%%%%%%%%%%%%%%%%%%%%%%%%%%%%%%    
\begin{figure}[t]
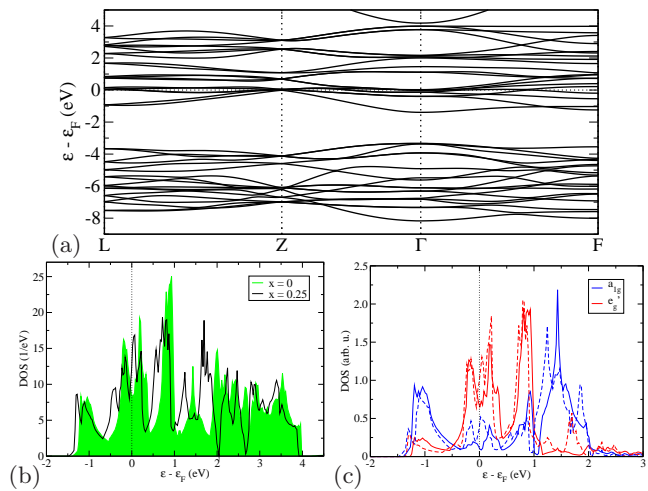

\centering
(a)\hspace*{-0.3cm}\includegraphics*[height=3.25cm]{Figures/bdstruc.eps}
(b)\hspace*{-0.3cm}\includegraphics*[height=2.95cm]{Figures/totdos.eps}
(c)\hspace*{-0.3cm}\includegraphics*[height=2.95cm]{Figures/partdos.eps}
\caption{(Color online) LDA spectral data. (a) Band structure of undoped V$_2$O$_3$ along
high symmetry lines. (b) Total density of states (DOS) of the doped supercell calculations for
different values of Cr doping. (c) Partial DOS projected onto the
$a_{1g}$ and $e_{g}'$ basis orbitals of one
vanadium atom.\label{fig:totdos_supercell}}
\end{figure}
%%%%%%%%%%%%%%%%%%%%%%%%%%%%%%%%%%%%%%%%%%%%%%%%%%%%%%%%%%%%%%%%%%%%%%%%%%%%%%% 
The results concerning the spectral properties are summarized in
Fig.~\ref{fig:totdos_supercell}. For undoped V$_2$O$_3$, the Kohn-Sham (KS) band
structure shows the well-known transition-metal-oxide splitting into completely filled 
oxygen-like bands (from about -8~eV to -4~eV) and partially filled vanadium $3d$-like
bands  (around the Fermi energy $\varepsilon_{\rm F}$). The latter show the above-mentioned 
ligand-field splitting into a partially filled $t_{2g}$ and an empty $e_g$ manifold. 
The structural relaxation within LDA leads to a less pronounced gap between these two compared 
to previous work using experimental values for the Wyckoff positions.~\cite{mat94,gri12}
The addition of one electron via one Cr impurity atom does not lead to 
significant changes of the KS density of states (DOS), apart from a seemingly small overall 
generic shift in energy. The trigonal crystal-field splitting of the $t_{2g}$ manifold into two
 $e_g'$ and one $a_{1g}$ state is also qualitatively very similar for all doping levels. 
Especially the $a_{1g}$ state keeps its pronounced bonding-antibonding structure caused
by the relatively large hopping in the crystallographic $c$~direction. Note
further that in the doped supercell calculations (with more than four V/Cr atoms), the $e_g'$ states 
are formally nondegenerate. Descriptively speaking, they can be distingiushed whether or not they 
point towards a Cr atom. The respective occupation values are averaged in the following, if
necessary.

\begin{figure}[t]
(a)\hspace*{-0.3cm}\includegraphics*[height=2.7cm]{Figures/fatbands_undoped.eps}
\hspace*{-0.1cm}
(b)\hspace*{-0.1cm}\includegraphics*[height=2.7cm]{Figures/fatbands_x0.25.eps}\\
(c)\hspace*{-0.3cm}\includegraphics*[height=2.7cm]{Figures/bdstrucmorezoom_undoped.eps}
\hspace*{-0.1cm}
(d)\hspace*{-0.1cm}\includegraphics*[height=2.7cm]{Figures/bdstrucmorezoom_x0.25.eps}\\
\caption{Low-energy KS band structure for (a,c) undoped V$_2$O$_3$ and
(b,d) Cr-doped with $x=0.25$. (a,b) Color-resolved $a_{1g}$ (blue) and $e_g'$ (red) weight. 
The respective light colours (cyan/pink) refer to the corresponding Cr-based $t_{2g}$ 
orbitals.\label{fig:lda_lowenergybdstruc}}
\includegraphics*[width=0.95\linewidth]{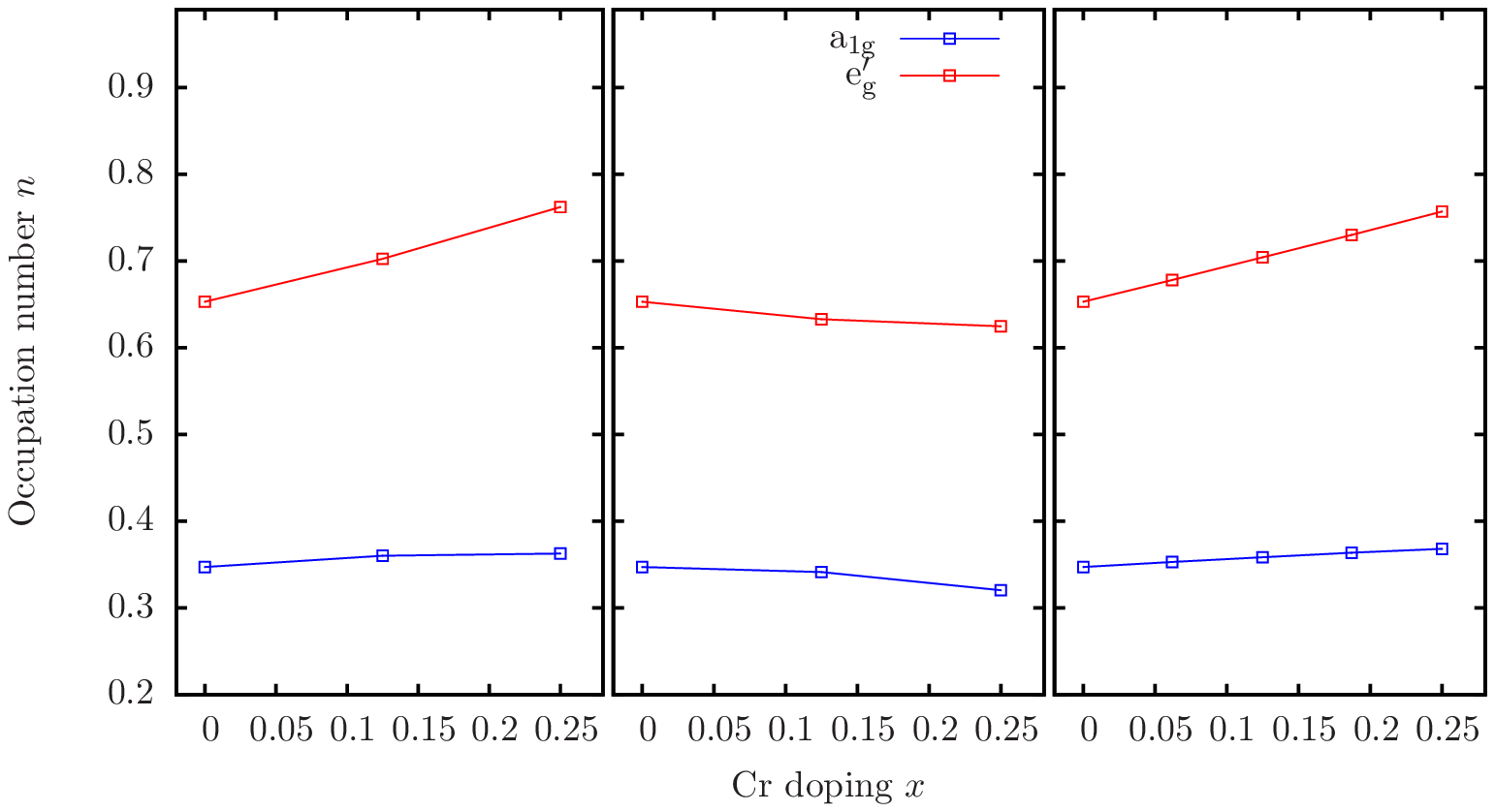}
\caption{(Color online) Average LDA occupation of the $a_{1g}$ and $e_g'$ orbitals with
Cr doping. Left: from supercell calculations, averaged over Cr and V,
middle: from supercell calculations, averaged over V only and right: within VCA.
\label{fig:occ_lda_withandwoV_supercell}}
\end{figure}

Fig.~\ref{fig:lda_lowenergybdstruc} displays the LDA-derived KS band
structure near the Fermi energy with and without doping together with the $k$-resolved
respective $a_{1g}$, $e_g'$ weight on each band (via so-called ``fatbands''). 
The latter again reflect the $a_{1g}$ bonding/antibonding character and the more
localized $e_g'$ behavior associated with narrower bands near the Fermi level. 
Notably however, small $a_{1g}$ spectral-weight contribution can also be found near 
$\varepsilon_{\rm F}$. In the undoped case the $Z$ point in the first Brillouin zone (BZ) 
marks close to the Fermi level a manifold band-crossing site, while at $\Gamma$ unoccupied
electron-like bands barely touch $\varepsilon_{\rm F}$. On the contrary for 
V$_{1.5}$Cr$_{0.5}$O$_2$ the multiple band crossings at $Z$ have disappeared and the 
electron-like bands at $\Gamma$ have now gained substantial occupation. It thus seems
as if on the LDA level the Cr impurity resolves some delicate features of the V$_2$O$_3$ 
fermiology and transfers the electronic structure to a more stabilized regime.
Interestingly, from Fig.~\ref{fig:lda_lowenergybdstruc}(a,c) it becomes evident that the empty 
$a_{1g}$-like band along $Z$-$\Gamma$ for $x$=0 crosses $\varepsilon_{\rm F}$ towards
$\Gamma$ for $x$=0.25. Note that in the latter doped case the occupied part of this band 
carries significant Cr-$a_{1g}$ weight. On the local level the LDA crystal-field splitting 
between $a_{1g}$ and $e_g'$ on Cr turns out to be slightly larger than on V.

%%%%%%%%%%%%%%%%%%%%%%%%%%%%%%%%%%%%%%%%%%%%%%%%%%%%%%%%%%%%%%%%%%%%%%%%%%%%%%%
%%%%%%%%%%%%%%%%%%%%%%%%%%%%%%%%%%%%%%%%%%%%%%%%%%%%%%%%%%%%%%%%%%%%%%%%%%%%%%%

As shown in the occupation-number plot of
Fig.~\ref{fig:occ_lda_withandwoV_supercell}, there is a polarization 
$\zeta$=$n_{e_g'}/n_{a_{1g}}$ of the $e_g'$ orbital against the $a_{1g}$ orbital 
already in stoichiometric V$_2$O$_3$ revealed by LDA calculation with eight transition-metal
ions in the primitive cell.
More interestingly, with Cr doping the average polarization $\zeta$ slightly increases.
This however mainly happens due to the increased local $\zeta$ on the Cr ion.
Fig.~\ref{fig:occ_lda_withandwoV_supercell} shows that the polarization is nearly
unchanged on the remaining V ions in the system with doping. Thus the additional electron 
due to Cr stays on its host ion and preferably occupies there the $e_g'$ states
in the LDA description. Note that the V occupation even seems to decrease slightly 
with Cr doping.

\section{Effect of electronic correlations within DFT+DMFT}

\subsection{Isostructural effects}

We now turn to an extended treatment of the electronic structure
by including realistic electron correlation within the DFT+DMFT
scope. To begin with, the structural parameters (including the LDA-relaxed Wyckoff positions) 
are not altered when turning to the advanced many-body investigation. This allows to 
clearly seperate electronic from structural influences of chromium doping.

%%%%%%%%%%%%%%%%%%%%%%%%%%%%%%%%%%%%%%%%%%%%%%%%%%%%%%%%%%%%%%%%%%%%%%%%%%%%%%%
\begin{figure}[b]
\centering
\parbox[c]{5cm}{(a)\includegraphics*[width=5cm]{Figures/dos_for_experiment.eps}}\\
\parbox[c]{8.5cm}{(b)\hspace*{-0.3cm}
\includegraphics*[width=3.85cm]{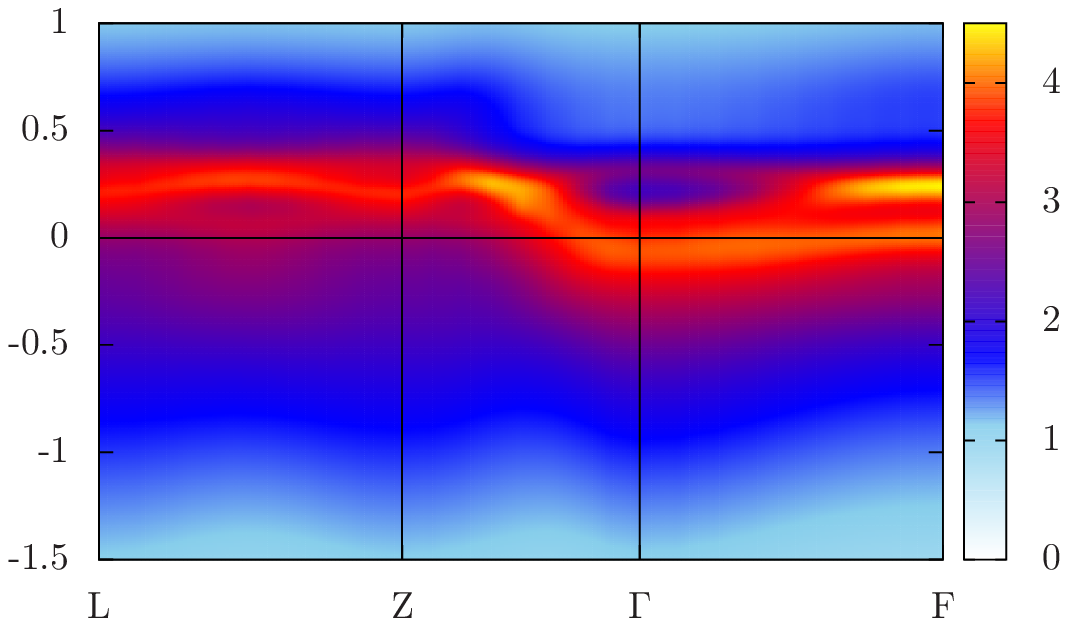}
(c)\hspace*{-0.3cm}
\includegraphics*[width=3.85cm]{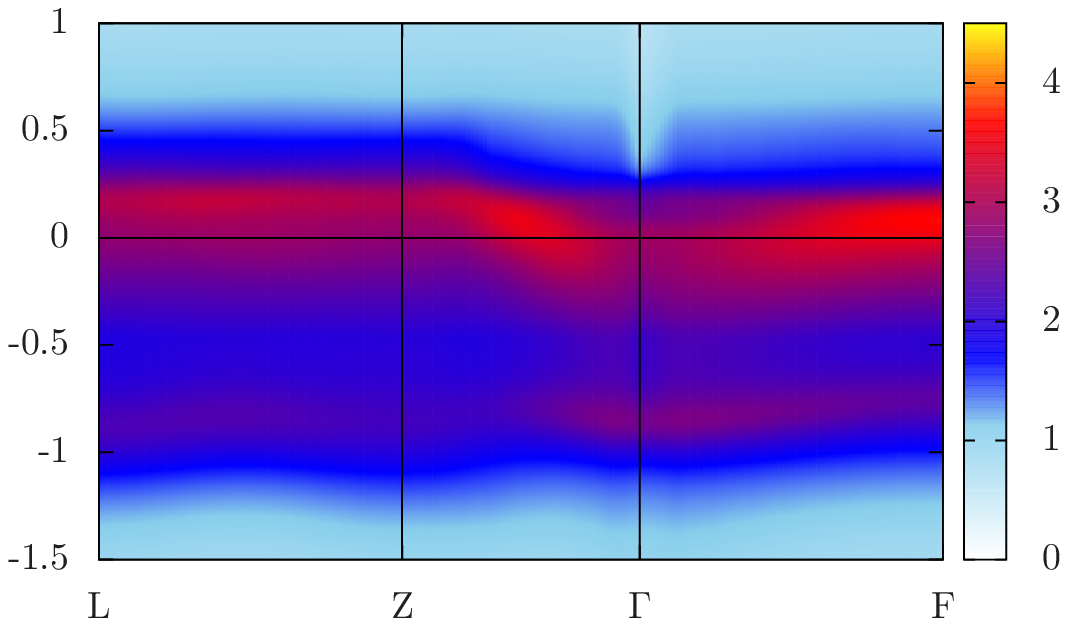}}
\caption{(Color online) DFT+DMFT spectral data at $T$=166K. 
(a) k-integrated total spectral function. Low-energy angle-resolved spectral function 
along high-symmetry lines (b) without and (c) with $x$=0.25 Cr doping.
\label{fig:maxent_spectra}}
\end{figure}
%%%%%%%%%%%%%%%%%%%%%%%%%%%%%%%%%%%%%%%%%%%%%%%%%%%%%%%%%%%%%%%%%%%%%%%%%%%%%%%

The obtained $k$-integrated spectral function above the antiferromagnetic
ordering temperature shown in Fig.~\ref{fig:maxent_spectra}a compares well to 
recent experimental data from photoemission spectroscopy by Mo {\sl et al.}~\cite{mo06}
In the stoichiometric case a low-energy resonance close to $-0.3$eV and a
shallow lower Hubbard peak at around $-1.5$eV appears in the spectrum. Although our
doping is much larger than in experiment, the qualitative feature due to Cr impurities,
namely a surpression of the low-energy quasiparticle (QP) peak with spectral-weight transfer to 
deeper more negative energies, agrees well with experimental findings. Note that from our 
calculation a {\sl sole} substitutional replacement of V by Cr at fixed crystal structure 
is not sufficient to render the system Mott-insulating. In other words, a Mott-insulating Cr-doped phase at 
the given LDA-optimized structural data cannot be stabilized without altering further parameters.
An intricate interplay between electron doping {\sl and} structural changes appears to trigger
the metal-to-insulator transition in V$_2$O$_3$.
Fig.~\ref{fig:maxent_spectra}(b,c) for the angle-resolved spectral data exhibits besides
again the spectral weight transfer, band narrowing and substantial incoherency effects on
the former noninteracting KS states well below room temperature. Interestingly especially
in the undoped case the QP signature is stronger in the perpendicular-to-$k_z$ plane through 
$\Gamma$ than the one through $Z$. The resonance at low-energy is identified
as a narrow QP state in the former plane. Also the renormalized band crossing along $Z$-$\Gamma$ 
is a key feature of the angle-resolved data.
%%%%%%%%%%%%%%%%%%%%%%%%%%%%%%%%%%%%%%%%%%%%%%%%%%%%%%%%%%%%%%%%%%%%%%%%%%%%%%%
\begin{figure}[b]
\centering
\includegraphics*[width=0.95\linewidth]{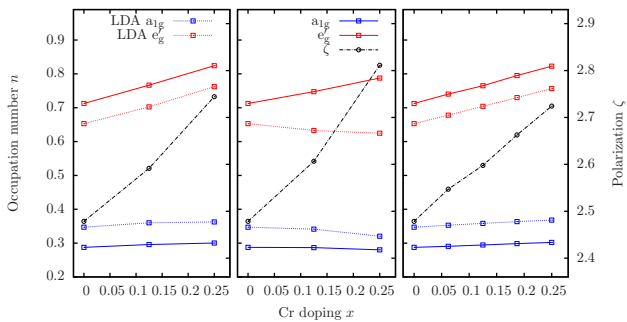}
\caption{(Color online) DFT+DMFT average occupation of the $a_{1g}$ and $e_g'$
  orbitals and orbital polarization $\zeta.$
  LDA data (dotted) repeated for comparison. Left: supercell calculations, 
  averaged over Cr and V, middle: supercell calculations, averaged over V only, 
  and right: VCA.
\label{fig:occ_csc_3}}
\end{figure}
%%%%%%%%%%%%%%%%%%%%%%%%%%%%%%%%%%%%%%%%%%%%%%%%%%%%%%%%%%%%%%%%%%%%%%%%%%%%%%%

In order to investigate in more detail how explicit Cr doping influences the
$n_{e_g'}/n_{a_{1g}}$ orbital polarization $\zeta$ in the correlated regime, we plot in 
Fig.~\ref{fig:occ_csc_3} the respective $t_{2g}$ occupations obtained
within the DFT+DMFT supercell calculations. From averaging over all sites (including
the Cr site) it is seen that a further increase of $\zeta$ compared to the LDA increase
with doping occurs. The doping-induced polarization increase is mainly based on the
the stronger $e_g'$ occupation with Cr. Note that our polarization values turn out to be in
good agreement with the numbers for Cr-doped V$_2$O$_3$ reported experimentally by 
Park {\sl et al.}.~\cite{par00} Although the occupation values averaged over {\sl all} 
transition-metal sites are most appealing for an experimental comparison, the physical 
implications can more easily been studied if only vanadium sites are considered. Interestingly, the above-mentioned $e_g'$ filling
and the following increase of $\zeta$ is even stronger in that case. Hence this proves 
that electronic correlations are readily responsible for the additional (dopant) electron 
to leave its ``home'' on Cr. The role of the Cr-originating electron is thus not limited 
to a pure spectator, but directly affects also the local electronic characteristics of V . 
From the difference of the summed occupations, one can estimate the part of electrons leaving
Cr to approximately 0.3-0.4. That the additional electron thus
increases the polarization can be seen as a direct consequence of the
spectrum, especially the strong bonding-antibonding character of
$a_{1g}$, leaving few (but not zero, as mentioned above) spectral weight near the Fermi energy. 
Figure~\ref{fig:occ_csc_3} shows furthermore that the orbital occupations (averaged
again over all transition-metal sites) obtained from VCA calculations for doped V$_2$O$_3$ are in 
excellent agreement with the supercell-related results. It furthermore stresses the linear 
character of the polarization increase with Cr doping.
%%%%%%%%%%%%%%%%%%%%%%%%%%%%%%%%%%%%%%%%%%%%%%%%%%%%%%%%%%%%%%%%%%%%%%%%%%%%%%%
\begin{figure}[t]
\centering
(a)\includegraphics*[width=7cm]{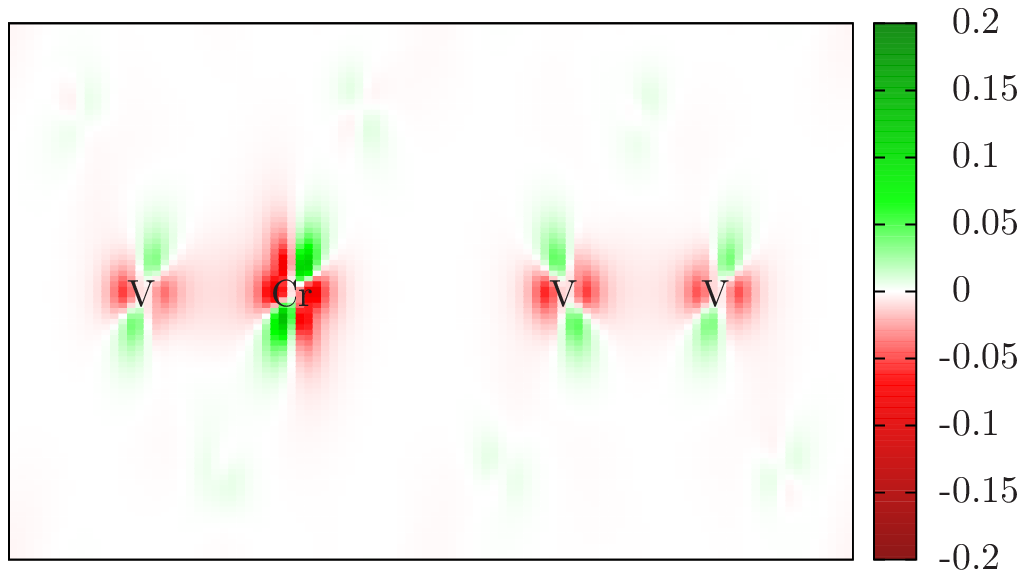}\\
(b)\includegraphics*[width=7cm]{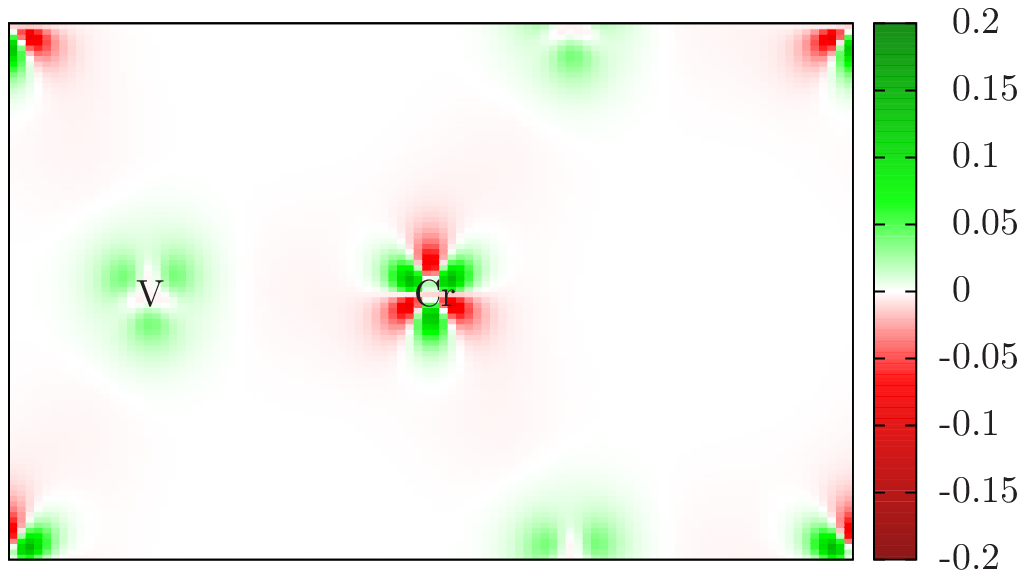}
\caption{(Color online) Difference $\Delta\rho$=$\rho_{\rm DFT+DMFT}$$-$$\rho_{\rm LDA}$
between correlated and LDA charge density for the doping level $x$=0.25.
(a) Along (here horizontally aligned) $c$-axis, (b) within the honeycomb-lattice $ab$-plane
of V$_2$O$_3$. 
\label{fig:chargedensity}}
\end{figure}
%%%%%%%%%%%%%%%%%%%%%%%%%%%%%%%%%%%%%%%%%%%%%%%%%%%%%%%%%%%%%%%%%%%%%%%%%%%%%%%

The correlation-induced local doping of V($e_g'$) via Cr can also be
visualized by plotting the charge-density difference $\Delta\rho$
between results from DFT+DMFT and LDA calculations
(cf. Fig.~\ref{fig:chargedensity}). While along the $c$-axis
$\Delta\rho$ dominantly exhibits the generic charge transfer within
the $t_{2g}$ manifold, the enhanced $e_g'$ filling on V is obvious
within the $ab$ honeycomb plane of the corundum structure.

An intuitive understanding for the correlation-induced charge transfer from
Cr to V is based on the increased local Coulomb energy introduced by
the stronger filled $t_{2g}$ of Cr compared to V. Nominally $t_{2g}$
half-filled, Coulomb interactions on Cr give rise to a larger self-energy and 
increased repulsion energy. Reducing the charge inhomogenity via transferring
part of the Cr electrons to V, leads to an overall energy lowering (partly
through gain of kinetic energy).

\subsection{Influence of the ratio $\frac ca$\label{caratio}}
\begin{figure}[t]
\centering
\includegraphics*[width=0.95\linewidth]{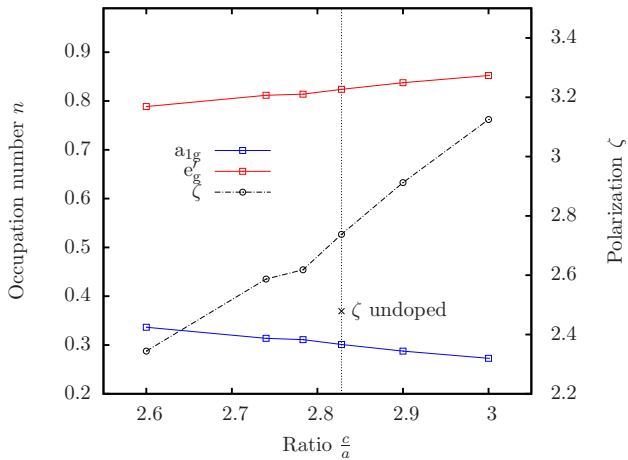}
\caption{(Color online) Occupation of the $a_{1g}$ and $e_g'$
  orbitals and orbital polarization $\zeta$ with varying ratio $\frac ca$ at 
  a doping of $x=0.25$. The vertical dotted line marks the experimentally 
  obtained value within the metallic phase used in the previous calculations.
\label{fig:occ_covera}}
\end{figure}

\begin{figure}[t]
\centering
\includegraphics*[width=0.95\linewidth]{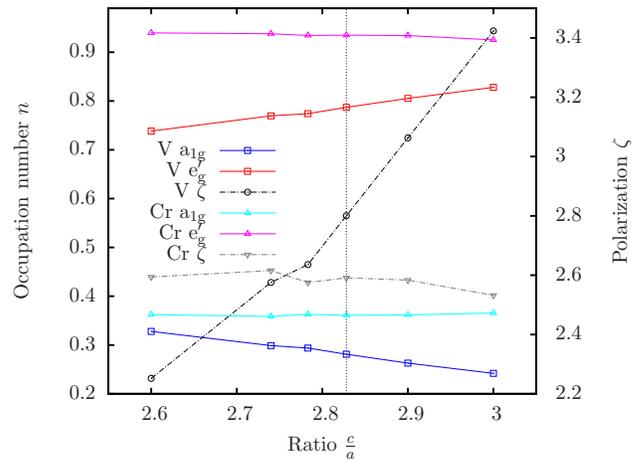}
\caption{(Color online) Vanadium and chromium resolved occupation numbers
  and $\zeta$ with varying ratio $\frac ca$ at a doping of
  $x=0.25$. Vertical dotted line as in Fig.~\ref{fig:occ_covera}.
  \label{fig:occ_covera_speciesresolved}}
\end{figure}

Apart from the explicit Cr doping, another important ingredient for
the exact behaviour of the orbital polarisation might be the ratio 
$\frac ca$ of the hexagonal crystallographic parameters. Since a small 
discontinuous jump thereof has been reported at the metal-insulator 
transition,~\cite{der70,rod11} it is worth investigating the
interplay of that ratio with explicit Cr doping. The value 
$\frac ca$$\sim$2.82 in the metallic regime is large compared to other 
materials with the same lattice structure,~\cite{der70} and is slightly 
lowered to about 2.78 in the Mott-insulating phase. Here we derive results
from supercell calculations with fixed unit-cell volume and LDA-relaxed
Wyckoff positions for the case of rather large Cr doping ($x$=0.25).
The overall evolution of the occupation numbers with $\frac ca$ ratio is 
displayed in Fig.~\ref{fig:occ_covera}. The decrease of the orbital 
polarization $\zeta$ with lower $\frac ca$ is the expected result. 
A more balanced value for the latter should render the transition-metal 
environment more isotropic and thus bring the orbital occupations closer
together. Similar behaviour has also been found when doing the same 
calculation for the undoped system (not shown here).
%%%%%%%%%%%%%%%%%%%%%%%%%%%%%%%%%%%%%%%%%%%%%%%%%%%%%%%%%%%%%%%%%%%%%%%%%%%%%%%
%%%%%%%%%%%%%%%%%%%%%%%%%%%%%%%%%%%%%%%%%%%%%%%%%%%%%%%%%%%%%%%%%%%%%%%%%%%%%%%
However, this does not imply that the increase of polarization due to
explicit Cr doping can completely be reversed due to the change of the
$\frac ca$ ratio. Through comparison to the polarization in the undoped case
it is obvious that the doping-driven enhancement of $\zeta$ may qualitatively 
still be strong enough to overcome the reduction via a decreased $\frac ca$ ratio
(see Fig.~\ref{fig:occ_covera}). As another observation, by studying the occupation 
numbers in an atomic-species-resolved way, shown in
Fig.~\ref{fig:occ_covera_speciesresolved}, it can be seen that the
variation of $\zeta$ is completely related to the
vanadium atoms. The chromium orbital occupations are essentially
unaffected by any change of $\frac ca$. A possible reason for that is
the initially higher crystal-field splitting of the respective Cr
orbitals. Consequently, at least the Cr-orbital-related part of the
doping-related polarization increase will persist when the $\frac ca$
ratio changes in the Mott-insulating phase.
%%%%%%%%%%%%%%%%%%%%%%%%%%%%%%%%%%%%%%%%%%%%%%%%%%%%%%%%%%%%%%%%%%%%%%%%%%%%%%%
%%%%%%%%%%%%%%%%%%%%%%%%%%%%%%%%%%%%%%%%%%%%%%%%%%%%%%%%%%%%%%%%%%%%%%%%%%%%%%%

The physical origin of the discontinuous jump of the $\frac ca$ ratio
was previously shown to be primarily due to the increase of the unit-cell 
volume,~\cite{gri12} i.~e. was part of the negative-pressure effect. 
However our calculations show that DFT+DMFT relaxed values of $\frac ca$ in 
the metallic Cr-doped material at {\sl constant} unit-cell volume are found to 
be even higher than in the undoped case. Thus an {\sl increased} unit-cell size 
in the doped regime would surely lead to an (overall) lowering of $\frac ca$, 
as also shown experimentally.~\cite{rod11} A complete quantitative 
first-principles account of the V$_2$O$_3$ MIT has therefore to address the 
delicate interplay between crystal-structure changes and Cr-doping in both
the metallic and the insulating regime.

\section{Summary and Discussion}

This work demonstrates that substitutional doping of V$_2$O$_3$ with Cr atoms (non-isovalent
with V) has an sophisticated impact on the detailed correlated electronic
structure that cannot be traced back to a pure (negative) pressure effect. 
Dispersion and filling of the QP bands (especially along $k_z$) change
substantially by introducing Cr to the system. The orbital polarization between the
relevant low-energy $e_g'$ and $a_{1g}$ states increases due to the electron
doping. Notably correlation-induced charge transfers from Cr to V occur that
are absent on the LDA level. This can explain why no occupation changes can 
be measured and calculated if only pressure is used to directly trigger the MIT.  
Thus including the true electron-doping physics does not require an increase of the 
effective crystal-field splitting of the {\sl undoped} compound, as obtained when the 
MIT is modelled by an increase of the Hubbard $U$.~\cite{pot07} It has to be noted that
the increase of orbital polarization displayed herein is nearly linear
with the doping level. The experimentally observed discontinous jump of the
polarization thus has to be explained by further non-continuous
processes occuring at this fascinating metal-to-insulator
transition, most surely through the intricate coupling to the lattice. 
Further correlated calculations with doping in the
insulating regime are necessary to reveal more details on this matter.  
Moreover it is known from experiment that the dopant atoms in V$_2$O$_3$ are not
homogeneously distributed in the crystal.~\cite{lup12} But modellings
thereof ask for much larger supercells with different impurity
configurations, which at present is beyond the scope of practicable DFT+DMFT (and
would also be extremely expensive already on the LDA level). It is
nonetheless expected that such advanced modellings would mainly
amplify the effects that are described in this work.

%\appendix
%\section{}

\begin{acknowledgments}
  We would like to thank M. Sandri and M. Fabrizio for helpful
  discussions. This work has been supported by the European Union,
  Seventh Framework Programme, under the project GO~FAST, grant
  agreement no.~280555 and the DFG-FOR1346 program. 
  Computations have been performed at the Juropa
  Cluster of the J\"ulich Supercomputing Centre (JSC) under project
  number hhh08.
\end{acknowledgments}

\bibliography{bibextra}

%Merlin.mbs v4.21 2009-07-09.
\begin{thebibliography}{10}%
\makeatletter
\providecommand \@ifxundefined [1]{%
 \ifx #1\undefined \expandafter \@firstoftwo
 \else \expandafter \@secondoftwo
\fi
}%
\providecommand \@ifnum [1]{%
 \ifnum #1\expandafter \@firstoftwo
 \else \expandafter \@secondoftwo
\fi
}%
\providecommand \enquote [1]{``#1''}%
\providecommand \bibnamefont  [1]{#1}%
\providecommand \bibfnamefont [1]{#1}%
\providecommand \citenamefont [1]{#1}%
\providecommand\href[0]{\@sanitize\@href}%
\providecommand\@href[1]{\endgroup\@@startlink{#1}\endgroup\@@href}%
\providecommand\@@href[1]{#1\@@endlink}%
\providecommand \@sanitize [0]{\begingroup\catcode`\&12\catcode`\#12\relax}%
\@ifxundefined \pdfoutput {\@firstoftwo}{%
 \@ifnum{\z@=\pdfoutput}{\@firstoftwo}{\@secondoftwo}%
}{%
 \providecommand\@@startlink[1]{\leavevmode\special{html:<a href="#1">}}%
 \providecommand\@@endlink[0]{\special{html:</a>}}%
}{%
 \providecommand\@@startlink[1]{%
  \leavevmode
  \pdfstartlink
   attr{/Border[0 0 1 ]/H/I/C[0 1 1]}%
   user{/Subtype/Link/A<</Type/Action/S/URI/URI(#1)>>}%
  \relax
 }%
 \providecommand\@@endlink[0]{\pdfendlink}%
}%
\providecommand \url  [0]{\begingroup\@sanitize \@url }%
\providecommand \@url [1]{\endgroup\@href {#1}{\urlprefix}}%
\providecommand \urlprefix [0]{URL }%
\providecommand \Eprint[0]{\href }%
\@ifxundefined \urlstyle {%
  \providecommand \doi [1]{doi:\discretionary{}{}{}#1}%
}{%
  \providecommand \doi [0]{doi:\discretionary{}{}{}\begingroup
  \urlstyle{rm}\Url }%
}%
\providecommand \doibase [0]{http://dx.doi.org/}%
\providecommand \Doi[1]{\href{\doibase#1}}%
\providecommand \bibAnnote [3]{%
  \BibitemShut{#1}%
  \begin{quotation}\noindent
    \textsc{Key:}\ #2\\\textsc{Annotation:}\ #3%
  \end{quotation}%
}%
\providecommand \bibAnnoteFile [2]{%
  \IfFileExists{#2}{\bibAnnote {#1} {#2} {\input{#2}}}{}%
}%
\providecommand \typeout [0]{\immediate \write \m@ne }%
\providecommand \selectlanguage [0]{\@gobble}%
\providecommand \bibinfo [0]{\@secondoftwo}%
\providecommand \bibfield [0]{\@secondoftwo}%
\providecommand \translation [1]{[#1]}%
\providecommand \BibitemOpen[0]{}%
\providecommand \bibitemStop [0]{}%
\providecommand \bibitemNoStop [0]{.\EOS\space}%
\providecommand \EOS [0]{\spacefactor3000\relax}%
\providecommand \BibitemShut [1]{\csname bibitem#1\endcsname}%
%</preamble>
\bibitem{cas78}%
  \BibitemOpen
  \bibfield{author}{%
  \bibinfo {author} {\bibfnamefont{C.}~\bibnamefont{Castellani}}, \bibinfo
  {author} {\bibfnamefont{C.~R.}\ \bibnamefont{Natoli}},\ and\ \bibinfo
  {author} {\bibfnamefont{J.}~\bibnamefont{Ranninger}},\ }%
  \bibfield{journal}{%
  \bibinfo {journal} {Phys. Rev. B}\ }%
  \textbf{\bibinfo {volume} {18}},\ \bibinfo {pages} {4945} (\bibinfo {year}
  {1978})%
  \bibAnnoteFile{NoStop}{cas78}%
\bibitem{mat94}%
  \BibitemOpen
  \bibfield{author}{%
  \bibinfo {author} {\bibfnamefont{L.~F.}\ \bibnamefont{Mattheiss}},\ }%
  \bibfield{journal}{%
  \bibinfo {journal} {J. Phys.: Condens. Matter}\ }%
  \textbf{\bibinfo {volume} {6}},\ \bibinfo {pages} {6477} (\bibinfo {year}
  {1994})%
  \bibAnnoteFile{NoStop}{mat94}%
\bibitem{ezh99}%
  \BibitemOpen
  \bibfield{author}{%
  \bibinfo {author} {\bibfnamefont{S.~Y.}\ \bibnamefont{Ezhov}}, \bibinfo
  {author} {\bibfnamefont{V.~I.}\ \bibnamefont{Anisimov}}, \bibinfo {author}
  {\bibfnamefont{D.~I.}\ \bibnamefont{Khomskii}},\ and\ \bibinfo {author}
  {\bibfnamefont{G.~A.}\ \bibnamefont{Sawatzky}},\ }%
  \bibfield{journal}{%
  \bibinfo {journal} {Phys. Rev. Lett.}\ }%
  \textbf{\bibinfo {volume} {83}},\ \bibinfo {pages} {4136} (\bibinfo {year}
  {1999})%
  \bibAnnoteFile{NoStop}{ezh99}%
\bibitem{elf03}%
  \BibitemOpen
  \bibfield{author}{%
  \bibinfo {author} {\bibfnamefont{I.~S.}\ \bibnamefont{Elfimov}}, \bibinfo
  {author} {\bibfnamefont{T.}~\bibnamefont{Saha-Dasgupta}},\ and\ \bibinfo
  {author} {\bibfnamefont{M.~A.}\ \bibnamefont{Korotin}},\ }%
  \bibfield{journal}{%
  \bibinfo {journal} {Phys. Rev. B}\ }%
  \textbf{\bibinfo {volume} {68}},\ \bibinfo {pages} {113105} (\bibinfo {year}
  {2003})%
  \bibAnnoteFile{NoStop}{elf03}%
\bibitem{eye05}%
  \BibitemOpen
  \bibfield{author}{%
  \bibinfo {author} {\bibfnamefont{V.}~\bibnamefont{Eyert}}, \bibinfo {author}
  {\bibfnamefont{U.}~\bibnamefont{Schwingenschl\"ogl}},\ and\ \bibinfo {author}
  {\bibfnamefont{U.}~\bibnamefont{Eckern}},\ }%
  \bibfield{journal}{%
  \bibinfo {journal} {Europhys. Lett.}\ }%
  \textbf{\bibinfo {volume} {70}},\ \bibinfo {pages} {782} (\bibinfo {year}
  {2005})%
  \bibAnnoteFile{NoStop}{eye05}%
\bibitem{hel01}%
  \BibitemOpen
  \bibfield{author}{%
  \bibinfo {author} {\bibfnamefont{K.}~\bibnamefont{Held}}, \bibinfo {author}
  {\bibfnamefont{G.}~\bibnamefont{Keller}}, \bibinfo {author}
  {\bibfnamefont{V.}~\bibnamefont{Eyert}}, \bibinfo {author}
  {\bibfnamefont{D.}~\bibnamefont{Vollhardt}},\ and\ \bibinfo {author}
  {\bibfnamefont{V.~I.}\ \bibnamefont{Anisimov}},\ }%
  \bibfield{journal}{%
  \bibinfo {journal} {Phys. Rev. Lett.}\ }%
  \textbf{\bibinfo {volume} {86}},\ \bibinfo {pages} {5345} (\bibinfo {year}
  {2001})%
  \bibAnnoteFile{NoStop}{hel01}%
\bibitem{kel04}%
  \BibitemOpen
  \bibfield{author}{%
  \bibinfo {author} {\bibfnamefont{G.}~\bibnamefont{Keller}}, \bibinfo {author}
  {\bibfnamefont{K.}~\bibnamefont{Held}}, \bibinfo {author}
  {\bibfnamefont{V.}~\bibnamefont{Eyert}}, \bibinfo {author}
  {\bibfnamefont{D.}~\bibnamefont{Vollhardt}},\ and\ \bibinfo {author}
  {\bibfnamefont{V.~I.}\ \bibnamefont{Anisimov}},\ }%
  \bibfield{journal}{%
  \bibinfo {journal} {Phys. Rev. B}\ }%
  \textbf{\bibinfo {volume} {70}},\ \bibinfo {pages} {205116} (\bibinfo {year}
  {2004})%
  \bibAnnoteFile{NoStop}{kel04}%
\bibitem{pot07}%
  \BibitemOpen
  \bibfield{author}{%
  \bibinfo {author} {\bibfnamefont{A.~I.}\ \bibnamefont{Poteryaev}}, \bibinfo
  {author} {\bibfnamefont{J.~M.}\ \bibnamefont{Tomczak}}, \bibinfo {author}
  {\bibfnamefont{S.}~\bibnamefont{Biermann}}, \bibinfo {author}
  {\bibfnamefont{A.}~\bibnamefont{Georges}}, \bibinfo {author}
  {\bibfnamefont{A.~I.}\ \bibnamefont{Lichtenstein}}, \bibinfo {author}
  {\bibfnamefont{A.~N.}\ \bibnamefont{Rubtsov}}, \bibinfo {author}
  {\bibfnamefont{T.}~\bibnamefont{Saha-Dasgupta}},\ and\ \bibinfo {author}
  {\bibfnamefont{O.~K.}\ \bibnamefont{Andersen}},\ }%
  \bibfield{journal}{%
  \bibinfo {journal} {Phys. Rev. B}\ }%
  \textbf{\bibinfo {volume} {76}},\ \bibinfo {pages} {085127} (\bibinfo {year}
  {2007})%
  \bibAnnoteFile{NoStop}{pot07}%
\bibitem{gri12}%
  \BibitemOpen
  \bibfield{author}{%
  \bibinfo {author} {\bibfnamefont{D.}~\bibnamefont{Grieger}}, \bibinfo
  {author} {\bibfnamefont{C.}~\bibnamefont{Piefke}}, \bibinfo {author}
  {\bibfnamefont{O.~E.}\ \bibnamefont{Peil}},\ and\ \bibinfo {author}
  {\bibfnamefont{F.}~\bibnamefont{Lechermann}},\ }%
  \bibfield{journal}{%
  \bibinfo {journal} {Phys. Rev. B}\ }%
  \textbf{\bibinfo {volume} {86}},\ \bibinfo {pages} {155121} (\bibinfo {year}
  {2012})%
  \bibAnnoteFile{NoStop}{gri12}%
\bibitem{san13}%
  \BibitemOpen
  \bibfield{author}{%
  \bibinfo {author} {\bibfnamefont{M.}~\bibnamefont{Sandri}}, \bibinfo {author}
  {\bibfnamefont{M.}~\bibnamefont{Capone}},\ and\ \bibinfo {author}
  {\bibfnamefont{M.}~\bibnamefont{Fabrizio}},\ }%
  \bibfield{journal}{%
  \bibinfo {journal} {Phys. Rev. B}\ }%
  \textbf{\bibinfo {volume} {87}},\ \bibinfo {pages} {205108} (\bibinfo {year}
  {2013})%
  \bibAnnoteFile{NoStop}{san13}%
\bibitem{guo14}%
  \BibitemOpen
  \bibfield{author}{%
  \bibinfo {author} {\bibfnamefont{Y.}~\bibnamefont{Guo}}, \bibinfo {author}
  {\bibfnamefont{S.~J.}\ \bibnamefont{Clark}},\ and\ \bibinfo {author}
  {\bibfnamefont{J.}~\bibnamefont{Robertson}},\ }%
  \bibfield{journal}{%
  \bibinfo {journal} {J. Chem. Phys.}\ }%
  \textbf{\bibinfo {volume} {140}},\ \bibinfo {eid} {054702} (\bibinfo {year}
  {2014})%
  \bibAnnoteFile{NoStop}{guo14}%
\bibitem{mcw71}%
  \BibitemOpen
  \bibfield{author}{%
  \bibinfo {author} {\bibfnamefont{D.~B.}\ \bibnamefont{McWhan}}, \bibinfo
  {author} {\bibfnamefont{J.~B.}\ \bibnamefont{Remeika}}, \bibinfo {author}
  {\bibfnamefont{T.~M.}\ \bibnamefont{Rice}}, \bibinfo {author}
  {\bibfnamefont{W.~F.}\ \bibnamefont{Brinkman}}, \bibinfo {author}
  {\bibfnamefont{J.~P.}\ \bibnamefont{Maita}},\ and\ \bibinfo {author}
  {\bibfnamefont{A.}~\bibnamefont{Menth}},\ }%
  \bibfield{journal}{%
  \bibinfo {journal} {Phys. Rev. Lett.}\ }%
  \textbf{\bibinfo {volume} {27}},\ \bibinfo {pages} {941} (\bibinfo {year}
  {1971})%
  \bibAnnoteFile{NoStop}{mcw71}%
\bibitem{mcw73}%
  \BibitemOpen
  \bibfield{author}{%
  \bibinfo {author} {\bibfnamefont{D.~B.}\ \bibnamefont{McWhan}}, \bibinfo
  {author} {\bibfnamefont{A.}~\bibnamefont{Menth}}, \bibinfo {author}
  {\bibfnamefont{J.~B.}\ \bibnamefont{Remeika}}, \bibinfo {author}
  {\bibfnamefont{T.~M.}\ \bibnamefont{Rice}},\ and\ \bibinfo {author}
  {\bibfnamefont{W.~F.}\ \bibnamefont{Brinkman}},\ }%
  \bibfield{journal}{%
  \bibinfo {journal} {Phys. Rev. B}\ }%
  \textbf{\bibinfo {volume} {7}},\ \bibinfo {pages} {1920} (\bibinfo {year}
  {1973})%
  \bibAnnoteFile{NoStop}{mcw73}%
\bibitem{din14}%
  \BibitemOpen
  \bibfield{author}{%
  \bibinfo {author} {\bibfnamefont{Y.}~\bibnamefont{Ding}}, \bibinfo {author}
  {\bibfnamefont{C.-C.}\ \bibnamefont{Chen}}, \bibinfo {author}
  {\bibfnamefont{Q.}~\bibnamefont{Zeng}}, \bibinfo {author}
  {\bibfnamefont{H.-S.}\ \bibnamefont{Kim}}, \bibinfo {author}
  {\bibfnamefont{M.~J.}\ \bibnamefont{Han}}, \bibinfo {author}
  {\bibfnamefont{M.}~\bibnamefont{Balasubramanian}}, \bibinfo {author}
  {\bibfnamefont{R.}~\bibnamefont{Gordon}}, \bibinfo {author}
  {\bibfnamefont{F.}~\bibnamefont{Li}}, \bibinfo {author}
  {\bibfnamefont{L.}~\bibnamefont{Bai}}, \bibinfo {author}
  {\bibfnamefont{D.}~\bibnamefont{Popov}}, \bibinfo {author}
  {\bibfnamefont{S.~M.}\ \bibnamefont{Heald}}, \bibinfo {author}
  {\bibfnamefont{T.}~\bibnamefont{Gog}}, \bibinfo {author}
  {\bibfnamefont{H.-k.}\ \bibnamefont{Mao}},\ and\ \bibinfo {author}
  {\bibfnamefont{M.}~\bibnamefont{van Veenendaal}},\ }%
  \bibfield{journal}{%
  \bibinfo {journal} {Phys. Rev. Lett.}\ }%
  \textbf{\bibinfo {volume} {112}},\ \bibinfo {pages} {056401} (\bibinfo {year}
  {2014})%
  \bibAnnoteFile{NoStop}{din14}%
\bibitem{bao93}%
  \BibitemOpen
  \bibfield{author}{%
  \bibinfo {author} {\bibfnamefont{W.}~\bibnamefont{Bao}}, \bibinfo {author}
  {\bibfnamefont{C.}~\bibnamefont{Broholm}}, \bibinfo {author}
  {\bibfnamefont{S.~A.}\ \bibnamefont{Carter}}, \bibinfo {author}
  {\bibfnamefont{T.~F.}\ \bibnamefont{Rosenbaum}}, \bibinfo {author}
  {\bibfnamefont{G.}~\bibnamefont{Aeppli}}, \bibinfo {author}
  {\bibfnamefont{S.~F.}\ \bibnamefont{Trevino}}, \bibinfo {author}
  {\bibfnamefont{P.}~\bibnamefont{Metcalf}}, \bibinfo {author}
  {\bibfnamefont{J.~M.}\ \bibnamefont{Honig}},\ and\ \bibinfo {author}
  {\bibfnamefont{J.}~\bibnamefont{Spalek}},\ }%
  \bibfield{journal}{%
  \bibinfo {journal} {Phys. Rev. Lett.}\ }%
  \textbf{\bibinfo {volume} {71}},\ \bibinfo {pages} {766} (\bibinfo {year}
  {1993})%
  \bibAnnoteFile{NoStop}{bao93}%
\bibitem{par00}%
  \BibitemOpen
  \bibfield{author}{%
  \bibinfo {author} {\bibfnamefont{J.-H.}\ \bibnamefont{Park}}, \bibinfo
  {author} {\bibfnamefont{L.~H.}\ \bibnamefont{Tjeng}}, \bibinfo {author}
  {\bibfnamefont{A.}~\bibnamefont{Tanaka}}, \bibinfo {author}
  {\bibfnamefont{J.~W.}\ \bibnamefont{Allen}}, \bibinfo {author}
  {\bibfnamefont{C.~T.}\ \bibnamefont{Chen}}, \bibinfo {author}
  {\bibfnamefont{P.}~\bibnamefont{Metcalf}}, \bibinfo {author}
  {\bibfnamefont{J.~M.}\ \bibnamefont{Honig}}, \bibinfo {author}
  {\bibfnamefont{F.~M.~F.}\ \bibnamefont{de~Groot}},\ and\ \bibinfo {author}
  {\bibfnamefont{G.~A.}\ \bibnamefont{Sawatzky}},\ }%
  \bibfield{journal}{%
  \bibinfo {journal} {Phys. Rev. B}\ }%
  \textbf{\bibinfo {volume} {61}},\ \bibinfo {pages} {11506} (\bibinfo {year}
  {2000})%
  \bibAnnoteFile{NoStop}{par00}%
\bibitem{rod10}%
  \BibitemOpen
  \bibfield{author}{%
  \bibinfo {author} {\bibfnamefont{F.}~\bibnamefont{Rodolakis}}, \bibinfo
  {author} {\bibfnamefont{P.}~\bibnamefont{Hansmann}}, \bibinfo {author}
  {\bibfnamefont{J.-P.}\ \bibnamefont{Rueff}}, \bibinfo {author}
  {\bibfnamefont{A.}~\bibnamefont{Toschi}}, \bibinfo {author}
  {\bibfnamefont{M.~W.}\ \bibnamefont{Haverkort}}, \bibinfo {author}
  {\bibfnamefont{G.}~\bibnamefont{Sangiovanni}}, \bibinfo {author}
  {\bibfnamefont{A.}~\bibnamefont{Tanaka}}, \bibinfo {author}
  {\bibfnamefont{T.}~\bibnamefont{Saha-Dasgupta}}, \bibinfo {author}
  {\bibfnamefont{O.~K.}\ \bibnamefont{Andersen}}, \bibinfo {author}
  {\bibfnamefont{K.}~\bibnamefont{Held}}, \bibinfo {author}
  {\bibfnamefont{M.}~\bibnamefont{Sikora}}, \bibinfo {author}
  {\bibfnamefont{I.}~\bibnamefont{Alliot}}, \bibinfo {author}
  {\bibfnamefont{J.-P.}\ \bibnamefont{Iti{\'e}}}, \bibinfo {author}
  {\bibfnamefont{F.}~\bibnamefont{Baudelet}}, \bibinfo {author}
  {\bibfnamefont{P.}~\bibnamefont{Wzietek}}, \bibinfo {author}
  {\bibfnamefont{P.}~\bibnamefont{Metcalf}},\ and\ \bibinfo {author}
  {\bibfnamefont{M.}~\bibnamefont{Marsi}},\ }%
  \bibfield{journal}{%
  \bibinfo {journal} {Phys. Rev. Lett.}\ }%
  \textbf{\bibinfo {volume} {104}},\ \bibinfo {pages} {047401} (\bibinfo {year}
  {2010})%
  \bibAnnoteFile{NoStop}{rod10}%
\bibitem{rod11}%
  \BibitemOpen
  \bibfield{author}{%
  \bibinfo {author} {\bibfnamefont{F.}~\bibnamefont{Rodolakis}}, \bibinfo
  {author} {\bibfnamefont{J.-P.}\ \bibnamefont{Rueff}}, \bibinfo {author}
  {\bibfnamefont{M.}~\bibnamefont{Sikora}}, \bibinfo {author}
  {\bibfnamefont{I.}~\bibnamefont{Alliot}}, \bibinfo {author}
  {\bibfnamefont{J.-P.}\ \bibnamefont{Iti{\'e}}}, \bibinfo {author}
  {\bibfnamefont{F.}~\bibnamefont{Baudelet}}, \bibinfo {author}
  {\bibfnamefont{S.}~\bibnamefont{Ravy}}, \bibinfo {author}
  {\bibfnamefont{P.}~\bibnamefont{Wzietek}}, \bibinfo {author}
  {\bibfnamefont{P.}~\bibnamefont{Hansmann}}, \bibinfo {author}
  {\bibfnamefont{A.}~\bibnamefont{Toschi}}, \bibinfo {author}
  {\bibfnamefont{M.~W.}\ \bibnamefont{Haverkort}}, \bibinfo {author}
  {\bibfnamefont{G.}~\bibnamefont{Sangiovanni}}, \bibinfo {author}
  {\bibfnamefont{K.}~\bibnamefont{Held}}, \bibinfo {author}
  {\bibfnamefont{P.}~\bibnamefont{Metcalf}},\ and\ \bibinfo {author}
  {\bibfnamefont{M.}~\bibnamefont{Marsi}},\ }%
  \bibfield{journal}{%
  \bibinfo {journal} {Phys. Rev. B}\ }%
  \textbf{\bibinfo {volume} {84}},\ \bibinfo {pages} {245113} (\bibinfo {year}
  {2011})%
  \bibAnnoteFile{NoStop}{rod11}%
\bibitem{mbpp_code}%
  \BibitemOpen
  \bibfield{author}{%
  \bibinfo {author} {\bibfnamefont{B.}~\bibnamefont{Meyer}}, \bibinfo {author}
  {\bibfnamefont{C.}~\bibnamefont{Els\"{a}sser}}, \bibinfo {author}
  {\bibfnamefont{F.}~\bibnamefont{Lechermann}},\ and\ \bibinfo {author}
  {\bibfnamefont{M.}~\bibnamefont{F\"{a}hnle}},\ }%
  \emph{\bibinfo {title} {FORTRAN 90 Program for Mixed-Basis-Pseudopotential
  Calculations for Crystals}},\ \bibinfo {organization} {Max-Planck-Institut
  f\"{u}r Metallforschung, Stuttgart} (\bibinfo {year} {unpublished})%
  \bibAnnoteFile{NoStop}{mbpp_code}%
\bibitem{wer06}%
  \BibitemOpen
  \bibfield{author}{%
  \bibinfo {author} {\bibfnamefont{P.}~\bibnamefont{Werner}}, \bibinfo {author}
  {\bibfnamefont{A.}~\bibnamefont{Comanac}}, \bibinfo {author}
  {\bibfnamefont{L.}~\bibnamefont{de' Medici}}, \bibinfo {author}
  {\bibfnamefont{M.}~\bibnamefont{Troyer}},\ and\ \bibinfo {author}
  {\bibfnamefont{A.~J.}\ \bibnamefont{Millis}},\ }%
  \bibfield{journal}{%
  \bibinfo {journal} {Phys. Rev. Lett.}\ }%
  \textbf{\bibinfo {volume} {97}},\ \bibinfo {pages} {076405} (\bibinfo {year}
  {2006})%
  \bibAnnoteFile{NoStop}{wer06}%
\bibitem{triqs_code}%
  \BibitemOpen
  \bibfield{author}{%
  \bibinfo {author} {\bibfnamefont{M.}~\bibnamefont{Ferrero}}\ and\ \bibinfo
  {author} {\bibfnamefont{O.}~\bibnamefont{Parcollet}},\ }%
  \enquote{\bibinfo {title} {{TRIQS}: a {T}oolbox for {R}esearch in
  {I}nteracting {Q}uantum {S}ystems},}\ \url{http://ipht.cea.fr/triqs}%
  \bibAnnoteFile{NoStop}{triqs_code}%
\bibitem{ama08}%
  \BibitemOpen
  \bibfield{author}{%
  \bibinfo {author} {\bibfnamefont{B.}~\bibnamefont{Amadon}}, \bibinfo {author}
  {\bibfnamefont{F.}~\bibnamefont{Lechermann}}, \bibinfo {author}
  {\bibfnamefont{A.}~\bibnamefont{Georges}}, \bibinfo {author}
  {\bibfnamefont{F.}~\bibnamefont{Jollet}}, \bibinfo {author}
  {\bibfnamefont{T.~O.}\ \bibnamefont{Wehling}},\ and\ \bibinfo {author}
  {\bibfnamefont{A.~I.}\ \bibnamefont{Lichtenstein}},\ }%
  \bibfield{journal}{%
  \bibinfo {journal} {Phys. Rev. B}\ }%
  \textbf{\bibinfo {volume} {77}},\ \bibinfo {pages} {205112} (\bibinfo {year}
  {2008})%
  \bibAnnoteFile{NoStop}{ama08}%
\bibitem{ani05}%
  \BibitemOpen
  \bibfield{author}{%
  \bibinfo {author} {\bibfnamefont{V.~I.}\ \bibnamefont{Anisimov}}, \bibinfo
  {author} {\bibfnamefont{D.~E.}\ \bibnamefont{Kondakov}}, \bibinfo {author}
  {\bibfnamefont{A.~V.}\ \bibnamefont{Kozhevnikov}}, \bibinfo {author}
  {\bibfnamefont{I.~A.}\ \bibnamefont{Nekrasov}}, \bibinfo {author}
  {\bibfnamefont{Z.~V.}\ \bibnamefont{Pchelkina}}, \bibinfo {author}
  {\bibfnamefont{J.~W.}\ \bibnamefont{Allen}}, \bibinfo {author}
  {\bibfnamefont{S.-K.}\ \bibnamefont{Mo}}, \bibinfo {author}
  {\bibfnamefont{H.-D.}\ \bibnamefont{Kim}}, \bibinfo {author}
  {\bibfnamefont{P.}~\bibnamefont{Metcalf}}, \bibinfo {author}
  {\bibfnamefont{S.}~\bibnamefont{Suga}}, \bibinfo {author}
  {\bibfnamefont{A.}~\bibnamefont{Sekiyama}}, \bibinfo {author}
  {\bibfnamefont{G.}~\bibnamefont{Keller}}, \bibinfo {author}
  {\bibfnamefont{I.}~\bibnamefont{Leonov}}, \bibinfo {author}
  {\bibfnamefont{X.}~\bibnamefont{Ren}},\ and\ \bibinfo {author}
  {\bibfnamefont{D.}~\bibnamefont{Vollhardt}},\ }%
  \bibfield{journal}{%
  \bibinfo {journal} {Phys. Rev. B}\ }%
  \textbf{\bibinfo {volume} {71}},\ \bibinfo {pages} {125119} (\bibinfo {year}
  {2005})%
  \bibAnnoteFile{NoStop}{ani05}%
\bibitem{kar11}%
  \BibitemOpen
  \bibfield{author}{%
  \bibinfo {author} {\bibfnamefont{M.}~\bibnamefont{Karolak}}, \bibinfo
  {author} {\bibfnamefont{T.~O.}\ \bibnamefont{Wehling}}, \bibinfo {author}
  {\bibfnamefont{F.}~\bibnamefont{Lechermann}},\ and\ \bibinfo {author}
  {\bibfnamefont{A.~I.}\ \bibnamefont{Lichtenstein}},\ }%
  \bibfield{journal}{%
  \bibinfo {journal} {J. Phys.: Condens. Matter}\ }%
  \textbf{\bibinfo {volume} {23}},\ \bibinfo {pages} {085601} (\bibinfo {year}
  {2011})%
  \bibAnnoteFile{NoStop}{kar11}%
\bibitem{der70}%
  \BibitemOpen
  \bibfield{author}{%
  \bibinfo {author} {\bibfnamefont{P.~D.}\ \bibnamefont{Dernier}},\ }%
  \bibfield{journal}{%
  \bibinfo {journal} {J. Phys. Chem. Solids}\ }%
  \textbf{\bibinfo {volume} {31}},\ \bibinfo {pages} {2569} (\bibinfo {year}
  {1970})%
  \bibAnnoteFile{NoStop}{der70}%
\bibitem{mo06}%
  \BibitemOpen
  \bibfield{author}{%
  \bibinfo {author} {\bibfnamefont{S.-K.}\ \bibnamefont{Mo}}, \bibinfo {author}
  {\bibfnamefont{H.-D.}\ \bibnamefont{Kim}}, \bibinfo {author}
  {\bibfnamefont{J.~D.}\ \bibnamefont{Denlinger}}, \bibinfo {author}
  {\bibfnamefont{J.~W.}\ \bibnamefont{Allen}}, \bibinfo {author}
  {\bibfnamefont{J.-H.}\ \bibnamefont{Park}}, \bibinfo {author}
  {\bibfnamefont{A.}~\bibnamefont{Sekiyama}}, \bibinfo {author}
  {\bibfnamefont{A.}~\bibnamefont{Yamasaki}}, \bibinfo {author}
  {\bibfnamefont{S.}~\bibnamefont{Suga}}, \bibinfo {author}
  {\bibfnamefont{Y.}~\bibnamefont{Saitoh}}, \bibinfo {author}
  {\bibfnamefont{T.}~\bibnamefont{Muro}},\ and\ \bibinfo {author}
  {\bibfnamefont{P.}~\bibnamefont{Metcalf}},\ }%
  \bibfield{journal}{%
  \bibinfo {journal} {Phys. Rev. B}\ }%
  \textbf{\bibinfo {volume} {74}},\ \bibinfo {pages} {165101} (\bibinfo {year}
  {2006})%
  \bibAnnoteFile{NoStop}{mo06}%
\bibitem{lup12}%
  \BibitemOpen
  \bibfield{author}{%
  \bibinfo {author} {\bibfnamefont{L.~B.}\ \bibnamefont{S.~Lupi}}, \bibinfo
  {author} {\bibfnamefont{B.}~\bibnamefont{Mansart}}, \bibinfo {author}
  {\bibfnamefont{A.}~\bibnamefont{Perucchi}}, \bibinfo {author}
  {\bibfnamefont{A.}~\bibnamefont{Barinov}}, \bibinfo {author}
  {\bibfnamefont{P.}~\bibnamefont{Dudin}}, \bibinfo {author}
  {\bibfnamefont{E.}~\bibnamefont{Papalazarou}}, \bibinfo {author}
  {\bibfnamefont{F.}~\bibnamefont{Rodolakis}}, \bibinfo {author}
  {\bibfnamefont{J.-P.}\ \bibnamefont{Rueff}}, \bibinfo {author}
  {\bibfnamefont{S.}~\bibnamefont{Ravy}}, \bibinfo {author}
  {\bibfnamefont{D.}~\bibnamefont{Nicoletti}}, \bibinfo {author}
  {\bibfnamefont{P.}~\bibnamefont{Postorino}}, \bibinfo {author}
  {\bibfnamefont{P.}~\bibnamefont{Hansmann}}, \bibinfo {author}
  {\bibfnamefont{N.}~\bibnamefont{Parragh}}, \bibinfo {author}
  {\bibfnamefont{A.}~\bibnamefont{Toschi}}, \bibinfo {author}
  {\bibfnamefont{T.}~\bibnamefont{Saha-Dasgupta}}, \bibinfo {author}
  {\bibfnamefont{O.~K.}\ \bibnamefont{Andersen}}, \bibinfo {author}
  {\bibfnamefont{G.}~\bibnamefont{Sangiovanni}}, \bibinfo {author}
  {\bibfnamefont{K.}~\bibnamefont{Held}},\ and\ \bibinfo {author}
  {\bibfnamefont{M.}~\bibnamefont{Marsi}},\ }%
  \bibfield{journal}{%
  \bibinfo {journal} {Nature Communications}\ }%
  \textbf{\bibinfo {volume} {1}},\ \bibinfo {pages} {105} (\bibinfo {year}
  {2010})%
  \bibAnnoteFile{NoStop}{lup12}%
\end{thebibliography}%

\end{document}